\documentclass[prb,twocolumn, superscriptaddress, showpacs]{revtex4}

\usepackage{graphicx} \usepackage{amssymb} \usepackage[usenames]{color}

\usepackage{amsmath}
\usepackage{xspace}

\newcommand{\ltsim}{\mbox{{\raisebox{-0.4ex}{$\stackrel{<}{{\scriptstyle\sim
}}
$}}}}

\begin{document}

\title{Exchange parameters of copper-based quasi-two-dimensional
Heisenberg magnets
measured using high magnetic fields and muon-spin rotation} 
\author{P.A. Goddard}
\affiliation{University of Oxford Department of Physics, 
The Clarendon Laboratory, Parks Road,
Oxford OX1~3PU, United Kingdom}
\author{John Singleton}
\affiliation{National High Magnetic Field Laboratory, 
Los Alamos National Laboratory, MS-E536, Los Alamos, NM 87545, USA}
\author{P. Sengupta}
\affiliation{Theoretical Division, Los Alamos National Laboratory, 
Los Alamos, NM 87545, USA}
\author{R.D. McDonald}
\affiliation{National High Magnetic Field Laboratory, 
Los Alamos National Laboratory, MS-E536, Los Alamos, NM 87545, USA}
\author{T.~Lancaster}
\affiliation{University of Oxford Department of Physics, 
The Clarendon Laboratory, Parks Road,
Oxford OX1~3PU, United Kingdom}
\author{S.J.~Blundell}
\affiliation{University of Oxford Department of Physics, 
The Clarendon Laboratory, Parks Road,
Oxford OX1~3PU, United Kingdom}
\author{F.L.~Pratt}
\affiliation{ISIS Facility, Rutherford Appleton Laboratory,
Chilton, Oxfordshire, OX11~0QX, United Kingdom}
\author{S.~Cox} 
\affiliation{National High Magnetic Field Laboratory, 
Los Alamos National Laboratory, MS-E536, Los Alamos, NM 87545, USA}
\author{N. Harrison} 
\affiliation{National High Magnetic Field Laboratory, 
Los Alamos National Laboratory, MS-E536, Los Alamos, NM 87545, USA}
\author{J.L. Manson}
\affiliation{Department of Chemistry and Biochemistry, 
Eastern Washington University, Cheney, WA 99004, USA}
\author{H.I. Southerland}
\affiliation{Department of Chemistry and Biochemistry, 
Eastern Washington University, Cheney, WA 99004, USA}
\author{J.A. Schlueter}
\affiliation{Materials Science Division, 
Argonne National Laboratory, Argonne, IL 60439, USA}

\begin{abstract}
Pulsed-field magnetization experiments
(fields $B$ of up to 85~T
and temperatures $T$ down to 0.4~K)
are reported on nine organic
Cu-based two-dimensional (2D) 
Heisenberg magnets. 
All compounds show a low-$T$
magnetization that is concave as a function of $B$,
with a sharp ``elbow'' transition
to a constant value at a field $B_{\rm c}$.
Monte-Carlo simulations including a finite
interlayer exchange energy $J_{\perp}$
quantitatively reproduce the 
data; the concavity indicates the effective dimensionality 
and $B_{\rm c}$ is an accurate measure of the
in-plane exchange energy $J$.
Using these values and Ne\'el temperatures
measured by muon-spin rotation, it is also possible
to obtain a quantitative estimate of $|J_{\perp}/J|$. 
In the light of these results,
it is suggested that in magnets of the 
form [Cu(HF$_2$)(pyz)$_2$]X,
where X is an anion,
the sizes of $J$ and $J_{\perp}$ are controlled
by the tilting of the pyrazine (pyz) 
molecule with respect to the 2D planes.
\end{abstract}


\maketitle
\section{Introduction}
Systems that can be described 
by the $S=\frac{1}{2}$ two-dimensional (2D)
square-lattice quantum Heisenberg antiferromagnet 
model~\cite{q2dreview,schollwock,manousakis} 
continue to attract
considerable experimental~\cite{christensen,vajk}
and theoretical~\cite{deng,beard,sengupta,zhang} attention.
Recent impetus
has been added to this field by
suggestions that antiferromagnetic fluctuations
from $S=\frac{1}{2}$ ions on a square lattice 
play a pivotal role in the mechanisms for
superconductivity in the ``high $T_{\rm c}$'' 
cuprates~\cite{manousakis,schrieffer,dai,stock,julian,harrison}
and other correlated-electron systems~\cite{pines};
moreover, 2D Heisenberg magnets have been suggested
as possible test-beds for processes applicable to
quantum computation~\cite{christensen,zhang}.

Though long-range magnetic order cannot occur above $T=0$
in a true 2D Heisenberg system~\cite{schollwock,mermin}, 
real materials that contain layers
approximating to 2D 
Heisenberg systems~\cite{christensen,schollwock,choi,mansonchemcomm} 
invariably possess
interlayer coupling that can lead to a finite 
Ne\'el temperature~\cite{christensen,choi,lancasterprb}.
In this context, 
synthesis of organic complexes
containing ions such as Cu$^{2+}$, 
neutral bridging ligands~\cite{christensen}
and coordinating anion molecules~\cite{choi,mansonchemcomm} has proved 
fruitful in the production of
a variety of one- and 2D 
magnetic systems~\cite{choi,deumel,lancaster1D,blundell}.
The current paper describes high-field 
magnetization measurements on
nine Cu-based quasi-2D Heisenberg magnets that employ
pyrazine (pyz) as a neutral bridging ligand~\cite{choi,mansonchemcomm,growth}. 
The data show that the 
field ($B$)-dependent, low-temperature ($T$)
magnetization $M(B)$ shows a characteristic 
sharp ``elbow'' feature at the
transition to saturation, with a 
concave curvature at lower $B$.
Monte-Carlo evaluations of a 2D Heisenberg square lattice
with an additional interlayer exchange
coupling energy $J_{\perp}$
reproduce the 
data quantitatively; the degree of
concavity depends on the effective 
dimensionality of the system,
whilst the field at which the ``elbow'' occurs 
is an accurate measure of the
in-plane exchange energy $J$. 
Using these $J$ values in conjunction with
Ne\'el temperatures deduced from muon-spin rotation
($\mu$SR), 
it is then possible
to gain a good estimate of the exchange
anisotropy $|J_{\perp}/J|$ for all of the magnets.

Having established these findings
using the whole range of compounds,
we suggest that in magnets of the
form [Cu(HF$_2$)(pyz)$_2$]X,~\cite{mansonchemcomm,lancasterprb} 
where X is an anion, $J$ and $J_{\perp}$
may be influenced by the tilting of the pyrazine 
(pyz) molecule with respect to the 2D planes.

\section{Experimental details}

\begin{figure}[htbp]
\centering
\includegraphics[width=7.0cm]{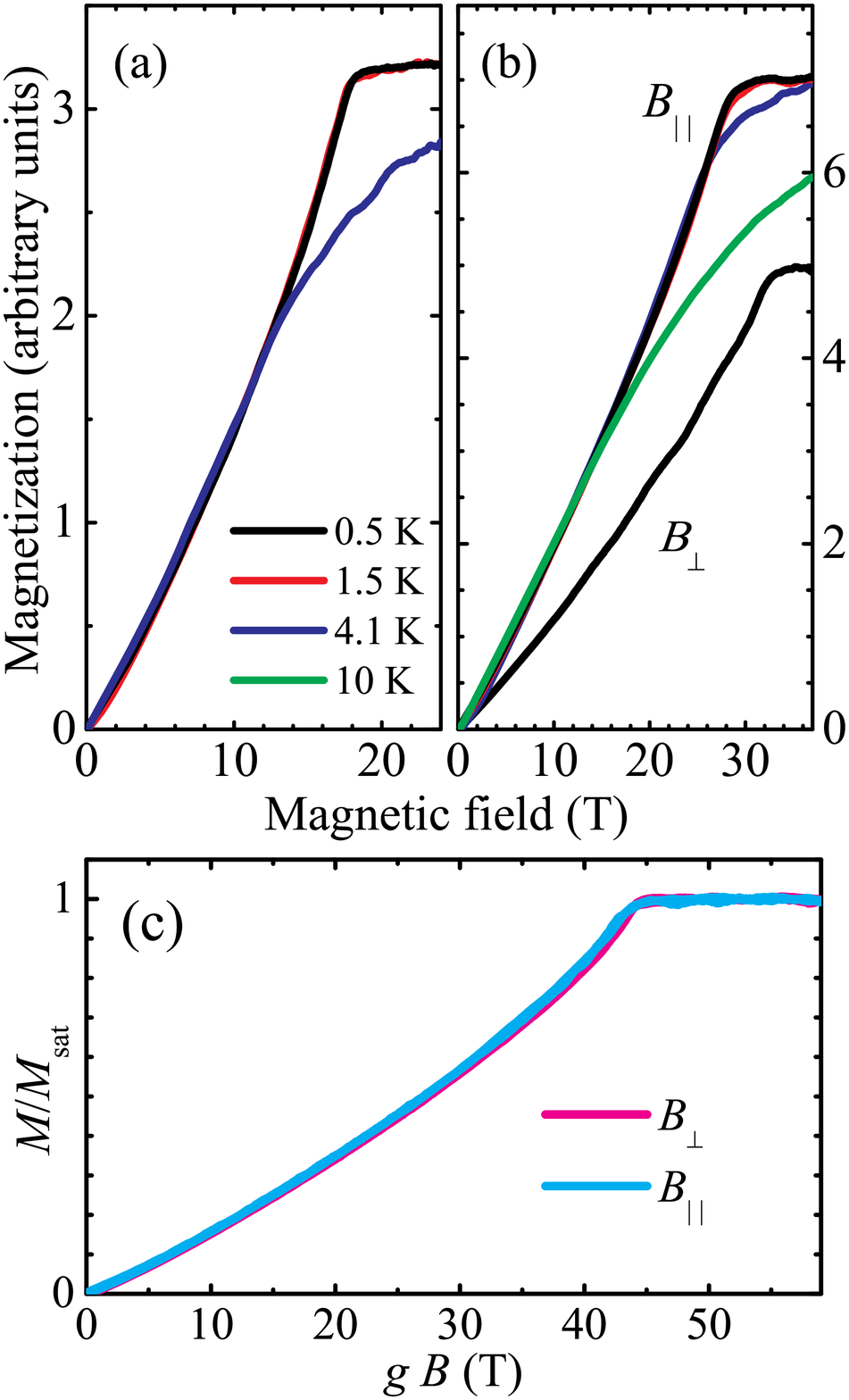}
\caption{(a)~Magnetization $M$ of
[Cu(HF$_2$)(pyz)$_2$](BF$_4$)
powder versus field $B$; data
for $T=0.5$, 1.5
and 4.1~K are shown (traces
for 0.5 and 1.5~K overlie).
(b)~Magnetization of
[CuF$_2$(pyz)](H$_2$O)$_2$ single crystals
with $B$ applied parallel 
(upper 4 traces) and perpendicular (lower trace, 0.5~K)
to the 2D planes. Data for $T=0.5$, 1.5, 4.1 and 10~K
are shown for the $B_{||}$ case.
The ``elbow'' denoting saturation
occurs at $B_{\rm c} = 28.8$~T for
$B_{||}$ and $B_{\rm c} = 33.1$~T for
$B_{\perp}$. (c)~Normalized $M$ data ($T=0.5$~K) for
[Cu(HF$_2$)(pyz)$_2$]ClO$_4$ single crystals versus
$gB$, where $g$ is the appropriate $g$-factor,
for $B$ parallel($B_{||}$) and perpendicular ($B_{\perp}$)
to the layers.}
\label{fig1}
\end{figure}

\begin{table*}
\centering
\begin{tabular}{|l|l|l|l|l|l|}
\hline
Compound & $B_{\rm c}$ (T) & $g$ & $J$~(K) & $T_{\rm N}$~(K) & $|J_{\perp}/J|$  \\ 
\hline
[Cu(HF$_2$)(pyz)$_2$]BF$_4$ & 18.0 & $2.13\pm 0.01$ & -6.3 & 1.54 & $9 \times 10^{-4}$ \\

[Cu(HF$_2$)(pyz)$_2$]ClO$_4$ $B_{\perp}$ & 19.1 & $2.30\pm 0.01$ & -7.3 & 1.94 & $2\times 10^{-3}$ \\

[Cu(HF$_2$)(pyz)$_2$]ClO$_4$ $B_{||}$ & 20.9 & $2.07 \pm 0.01$ & -7.2 & 1.94 & $2\times 10^{-3}$ \\

[Cu(HF$_2$)(pyz)$_2$]PF$_6$  & 35.5 & $2.11 \pm 0.01$ & -12.4 & 4.31 & $1 \times 10^{-2}$ \\

[Cu(HF$_2$)(pyz)$_2$]SbF$_6$ & 37.6 & $2.11\pm 0.01$ & -13.3 & 4.31 & $9\times 10^{-3}$ \\

[Cu(HF$_2$)(pyz)$_2$]AsF$_6$ & 36.1 & $2.16\pm 0.02$ & -12.9 & 4.34 & $1 \times 10^{-2}$ \\

[Cu(pyo)$_2$(pyz)$_2$](ClO$_4$)$_2$ $B_{\perp}$ & 20.8 & $2.30\pm 0.01$ & -7.9 & $<2.0$ & $\ltsim~ 10^{-3}$ \\

[Cu(pyo)$_2$(pyz)$_2$](ClO$_4$)$_2$ $B_{||}$ & 22.2 & $2.07\pm 0.01$ & -7.6 & $<2.0$ & $\ltsim~10^{-3}$ \\

CuF$_2$(pyz)(H$_2$O)$_2$ $B_{\perp}$ & 33.1 & $2.09\pm 0.02$ & -11.6 & 2.54 & $3\times 10^{-4}$ \\

CuF$_2$(pyz)(H$_2$O)$_2$ $B_{||}$ & 28.8 & $2.42\pm 0.01$ & -11.5 & 2.54 & $4\times 10^{-4}$ \\

Cu(pyz)$_2$(ReO$_4$)$_2$ & 42.7 & $2.10\pm 0.05$ & -14.9 & 4.2 & $3 \times 10^{-3}$ \\

Cu(pyz)$_2$(H$_2$O)$_2$Cr$_2$O$_7$ & 13.3 & $2.13 \pm 0.01$ & -4.7 & $<1.6$ & $\ltsim~ 1\times10^{-2}$ \\ 
\hline
\end{tabular}
\caption{The quasi-2D magnets studied in this work,
along with their saturation fields $B_{\rm c}$ and $g$-factors
(here pyz is pyrazine, pyo is pyridine-N-oxide).
Data for oriented single crystals are indicated by $B_{||}$
($B$ parallel to 2D layers) and $B_{\perp}$ ($B$
perpendicular to 2D layers); other data are for powders.
In the latter cases, an average $g$ was evaluated
from single-crystal EPR data using standard
formulae~\cite{abragam}.
The intralayer exchange energy is calculated using
Eq.~\ref{revdianpaisley};
typical uncertainties in the values of
$J$ resulting from uncertainties in $g$ and $B_{\rm c}$
are $\pm 0.1$~K.
Ne\'el temperatures $T_{\rm N}$ were measured 
to $\pm 0.04$~K using $\mu$SR~\cite{blundell},
apart from Cu(pyz)$_2$(ReO$_4$)$_2$, where the transition was observed
in heat capacity data~\cite{growth} (typical uncertainty $\pm 0.1$~K).
The anisotropy $|J_{\perp}/J|$ calculated using Eq.~\ref{anis}.}
\label{table1}
\end{table*}

The quasi-2D magnets studied in this
work are listed in Table~\ref{table1}.
The samples are produced in single or polycrystalline
form via aqueous chemical reaction between the
appropriate CuX$_2$ salts and stoichiometric
amounts of the ligands;
further details are given in
Refs.~\cite{mansonchemcomm,growth,entangle}, where
structural data derived from X-ray crystallography are also found.
Some compounds had crystals
large enough to permit measurements with a single, orientated
sample (Table~\ref{table1}).
In other cases, the materials
were either polycrystalline or the crystals
too small for accurate orientation; therefore
samples composed of many
randomly-orientated microcrystals, effectively powders, were used.
In addition to the characterization
described in Refs.\cite{mansonchemcomm,growth},
sample $g$-factors were measured~\cite{cox} using standard
electron paramagnetic resonance (EPR)~\cite{abragam}.

The pulsed-field $M$ experiments used a 1.5~mm bore, 
1.5~mm long, 1500-turn compensated-coil susceptometer,
constructed from 50-gauge high-purity copper wire~\cite{goddard,ho}. 
When a sample is within the coil,
the signal is $V \propto ({\rm d}M/{\rm d} t)$, where 
$t$ is the time. 
Numerical integration is used to evaluate $M$.~\cite{goddard}
The sample is mounted within a 1.3~mm diameter 
ampoule that can be moved in and out of the coil~\cite{goddard}.
Accurate values of $M$ are
obtained by subtracting
empty coil data from that measured
under identical conditions with the sample present.

Fields were provided by the
65~T short-pulse and 100~T multi-shot
magnets at NHMFL Los Alamos~\cite{boebinger} and a 60~T
short-pulse magnet at Oxford. 
The susceptometer was placed within a $^3$He 
cryostat providing $T$s
down to 0.4~K. $B$ was measured by 
integrating the voltage induced in a
ten-turn coil calibrated by observing the de Haas-van Alphen 
oscillations of the belly orbits
of the copper coils of the susceptometer~\cite{ho}.

In the cases that sufficient quantities of material
were available, Ne\'el temperatures $T_{\rm N}$ were measured using 
the zero-field muon-spin rotation ($\mu$SR) techniques 
described in Ref.~\cite{blundell}
(see also Refs.~\cite{lancasterprb,entangle});
the results are given in Table~\ref{table1}.
Ref.~\cite{blundell} also discusses the difficulties encountered when
using more conventional techniques to find $T_{\rm N}$
in quasi-2D magnets. 

\section{Experimental results}
Typical $M(H)$ data are shown in Fig.~\ref{fig1};
all compounds studied (Table~\ref{table1})
behaved in a very similar fashion.
At higher $T$, $M(B)$ is convex,
showing a gradual approach to saturation at high $B$. However,
as $T \rightarrow 0$, the $M(B)$ data become
concave at lower $B$, with a sharp, ``elbow''-like transition
to a constant saturation magnetization $M_{\rm sat}$ at higher $B$;
no further changes in $M$ occur to fields of 85~T with the current
materials.
We label the field at which the ``elbow'' occurs $B_{\rm c}$.
As shown in Fig.~\ref{fig1}b, $B_{\rm c}$ depends on
the crystal's orientation in the field.
However, in such cases,
the $M$ data become identical
to within experimental accuracy when 
plotted as $M/M_{\rm sat}$
versus $gB$, where $g$ is the
$g$-factor appropriate for that direction of $B$
(Fig.~\ref{fig1}(c) and Table~\ref{table1}).
This suggests that the $g$-factor anisotropy is responsible
for the observed angle dependence of $B_{\rm c}$.

\section{Monte-Carlo simulations of magnetization data}
The magnetic properties of the materials
in Table~\ref{table1} are well described by $S=\frac{1}{2}$
Cu$^{2+}$ spins on a square lattice. 
The layers are arranged in
a tetragonal structure~\cite{mansonchemcomm};
coupling between the layers
depends on the packing of the molecules between. 
To model data such as those in Fig.~\ref{fig1}, we assume
that the interaction
between the spins is purely Heisenberg-like~\cite{q2dreview},
resulting in the Hamiltonian
\begin{equation}
{\cal H} =J\sum_{\langle i,j \rangle_{xy}}{\bf S}_i\cdot {\bf S}_j +
J_\perp\sum_{\langle i,j \rangle_z}{\bf S}_i\cdot {\bf S}_j 
-h\sum_iS_i^z
\label{eq1}
\end{equation}
where $J$ ($J_\perp$) is the strength of the 
intra- (inter-) planar coupling
and $h=g\mu_{\rm B}B$ is the Zeeman energy
provided by the (uniform) magnetic induction $B$.
The first (second) summation refers to summing 
over all nearest neighbors
parallel (perpendicular) to the 2D $xy$-plane. 
The dependence of $M$
as a function
of $B$ is studied for a range of values of  $J_\perp /J$ 
[from  $J_\perp=0$ (completely decoupled layers) 
to $J_\perp/J=1$] using large-scale 
numerical simulations.
Note that the 
orientation of $B$
only affects $h$ through 
the anisotropy of the $g$-factor,
in agreement with the discussion of
Fig~\ref{fig1}(c) above.

The stochastic series expansion (SSE) method~\cite{sse1,sse2} is a
finite-$T$ Quantum Monte Carlo (QMC)
technique based on 
importance sampling of the diagonal matrix elements 
of the density matrix ${\rm e}^{-\beta H}$. 
Using the ``operator-loop''
cluster update~\cite{sse2}, the autocorrelation 
time for system sizes
considered here (up to $\approx 3\times 10^4$ spins) 
is at most a few Monte Carlo sweeps even 
at the critical $T$
for the onset of magnetic order~\cite{dloops}. 
Estimates of ground
state observables are obtained by using sufficiently 
large values of the
inverse $T$, $\beta$. We have further found 
that the statistics
of the data obtained can be significantly 
improved by the use of a tempering
scheme~\cite{tempering1,tempering2}. 
We use parallel tempering~\cite{tempering2}, 
where simulations are run simultaneously on a
parallel computer, using a fixed value of 
$J_{\perp}$ and different, but closely
spaced, values of $h$ over the entire 
range of fields up
to saturation. Along with the
usual Monte Carlo updates, we attempt 
to swap the values of fields for SSE
configurations (processes) with adjacent 
values of $h$ at regular intervals
(typically after every Monte Carlo step, 
each time attempting several
hundred swaps) according to a scheme that 
maintains detailed balance in the
space of the parallel simulations. 
This has favorable effects on the
simulation dynamics, and  reduces the
overall statistical errors (at the cost 
of introducing correlations
between the errors, of 
minor significance here). Implementation
of tempering schemes in the context of the 
SSE method is discussed
in Ref.~\cite{bow}.

\begin{figure}[htbp]
\centering
\includegraphics[width=7.5cm]{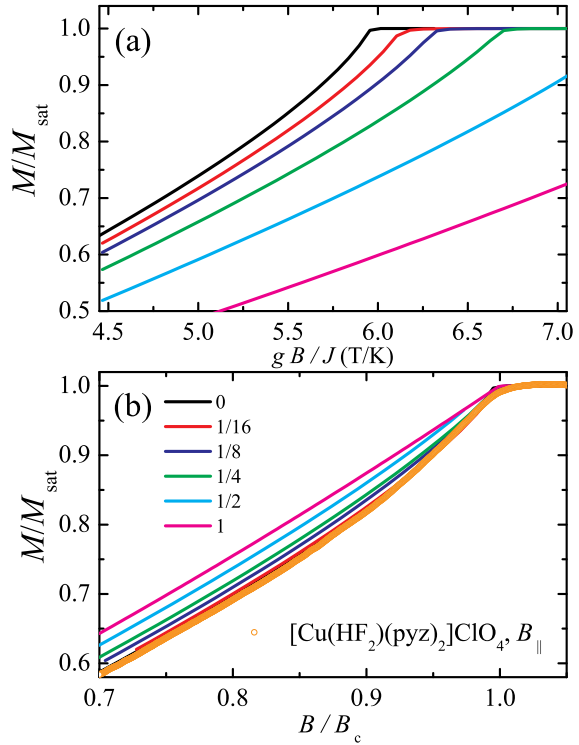}
\caption{(a)~Calculated 
magnetization $M$ of a spin-$\frac{1}{2}$
square lattice with added interlayer anisotropy
(see Eq.~\ref{eq1}).
Simulations are shown for $J_{\perp}/J = 0$ (uppermost curve)
$\frac{1}{16}$, $\frac{1}{8}$, $\frac{1}{4}$,
$\frac{1}{2}$ and $1$ (lowest curve).
Increasing $J_{\perp}/J$ raises the critical
field for saturation $B_{\rm c}$
and reduces the curvature of $M(B)$
below $B_{\rm c}$.
(b)~Comparison of the model shown in (a)
(curves) with experimental
data (points) for a [Cu(HF)$_2$(pyz)$_2$]ClO$_4$
crystal ($T=0.5$~K, field applied parallel to the 2D planes).
Both model and data are plotted in reduced units
$M/M_{\rm sat}$, $B/B_{\rm c}$. Note
that the experimental data lie between the
$J_{\perp}/J=0$ and $J_{\perp}/J = \frac{1}{16}$
model curves.}
\label{fig2}
\end{figure}
\section{Comparisons of model and data}
Fig.~\ref{fig2}(a) shows the predictions of
the model for low $T$, and 
Fig.~\ref{fig2}(b) shows a comparison
with typical experimental data.
This is made by plotting both model
results and experimental data in dimensionless 
units, $M/M_{\rm sat}$
and $B/B_{\rm c}$.
As $M_{\rm sat}$ is known, there is in effect only one fit parameter,
$B_{\rm c}$. The value of $B_{\rm c}$ 
is varied until there is a satisfactory
overlap of the data and one of the model curves~\cite{rosspoint}.
The curvature of the data and the presence of
the ``elbow'' place tight constraints on such a comparison;
that in Fig~\ref{fig2}(b) is typical of all of
the materials in Table~\ref{table1}, with their
$M(B)$ data
falling between, or closest to, the $J_{\perp}/J=0$
or $J_{\perp}/J=\frac{1}{16}$ numerical curves, 
indicating a high degree of anisotropy.
We shall give further justification for this
assertion below.

\begin{figure}[htbp]
\centering
\includegraphics[width=7.5cm]{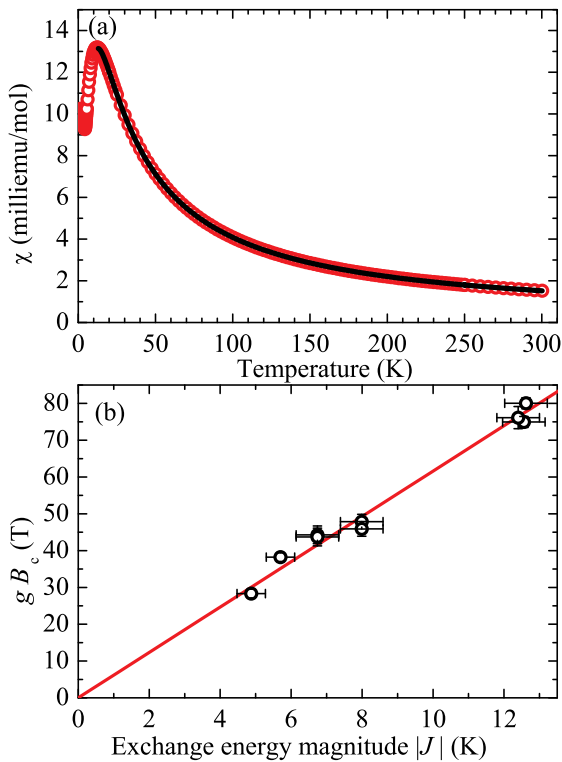}
\caption{(a)~Temperature dependence of the
low-field susceptibility $\chi$ of [Cu(HF)$_2$(pyz)$_2$]SbF$_6$
(points) fitted to the 2D Heisenberg expression (curve)
of Ref.~\cite{landee} to obtain an estimate of $J$.
(b)~Experimental
saturation fields times $g$-factor
($gB_{\rm c}$) plotted against 
values of $J$ deduced using fits such as
that in (a) (points). The straight-line
fit to the data yields 
$gB_{\rm c}/|J| = 6.2 \pm 0.2$~TK$^{-1}$.}
\label{fig3}
\end{figure}

For such highly anisotropic magnets, the model predicts
the ratio $gB_{\rm c}/|J|$ to take values in the range
$5.95~{\rm TK^{-1}}$ ($J_{\perp} =0$) to $6.10~{\rm TK^{-1}}$
($J_{\perp}/J = \frac{1}{16}$). Since the
experimental uncertainties involved in the location of $B_{\rm c}$
are $\sim 1-2\%$, and the errors in $g$ are $\sim 1\%$,
no significant loss in accuracy occurs
if we employ the mean value, 
\begin{equation}
\frac{gB_{\rm c}}{|J|} \approx 6.03~{\rm TK^{-1}},
\label{revdianpaisley}
\end{equation}
in what follows.
To check the model prediction, a fit of the $T$-dependent
low-field susceptibility following the method
of Ref.~\cite{landee} was used to determine 
$J$ independently for a selection 
of compounds (Fig.~\ref{fig3}(a))~\cite{footnote};
the values obtained are compared with $gB_{\rm c}$ in 
Fig~\ref{fig3}(b).
As can be seen, the points lie close to the line 
$gB_{\rm c}/|J|=6.2 \pm 0.2~{\rm TK^{-1}}$,
in good agreement with the predicted value (Eq.~\ref{revdianpaisley}).
As noted above, it is possible to determine the value of $B_{\rm c}$
to a very good accuracy ($\pm 1-2 \%$);
once the dimensionality of the magnet in question is seen
to fall within the limits $J_{\perp}/J \approx 0-\frac{1}{16}$
using fits such as those in Fig.~\ref{fig2}(b),
Eq.~\ref{revdianpaisley} 
almost certainly presents the
most accurate method for evaluating $J$.
Intralayer exchange energies
$J$ derived in this way from measured
values of $g$ and $B_{\rm c}$ are in Table~\ref{table1}.

Finally, an estimate of the anisotropy $|J_{\perp}/J|$ can be made
using the results of quantum-Monte-Carlo simulations of quasi-2D
Heisenberg antiferromagnets~\cite{yasuda}, which found
\begin{equation}
\frac{|J_{\perp}|}{|J|} = \exp\left(2.43-2.30\times\frac{|J|}{T_{\rm N}} \right),
\label{anis}
\end{equation}
where both $|J|$ and $T_{\rm N}$ are measured in K;
the resulting values are given in Table~\ref{table1}.
Note that Eq.~\ref{anis} is a rapidly-varying function of
$|J|/T_{\rm N}$; small shifts in either parameter
result in quite large changes in $|J_{\perp}/J|$.
Given the experimental and other errors in $J$
and $T_{\rm N}$ ($\sim$ a few $\%$), the 
derived values of $|J_{\perp}/J|$ will
probably be within a factor $\sim 2$ of the true values.
In spite of this {\it caveat}, the
$|J_{\perp}/J|$ values in Table~\ref{table1}
are all $\ltsim~0.01$, in good agreement with the
fits to the magnetization data (e.g. Fig~\ref{fig2})
that suggested $0~\ltsim ~|J_{\perp}/J| ~\ltsim ~\frac{1}{16}$
for all of the compounds.

\section{Systematic trends in the [Cu(HF$_2$)(pyz)$_2$]X family}
\begin{figure}[htbp]
\centering
\includegraphics[width=6cm]{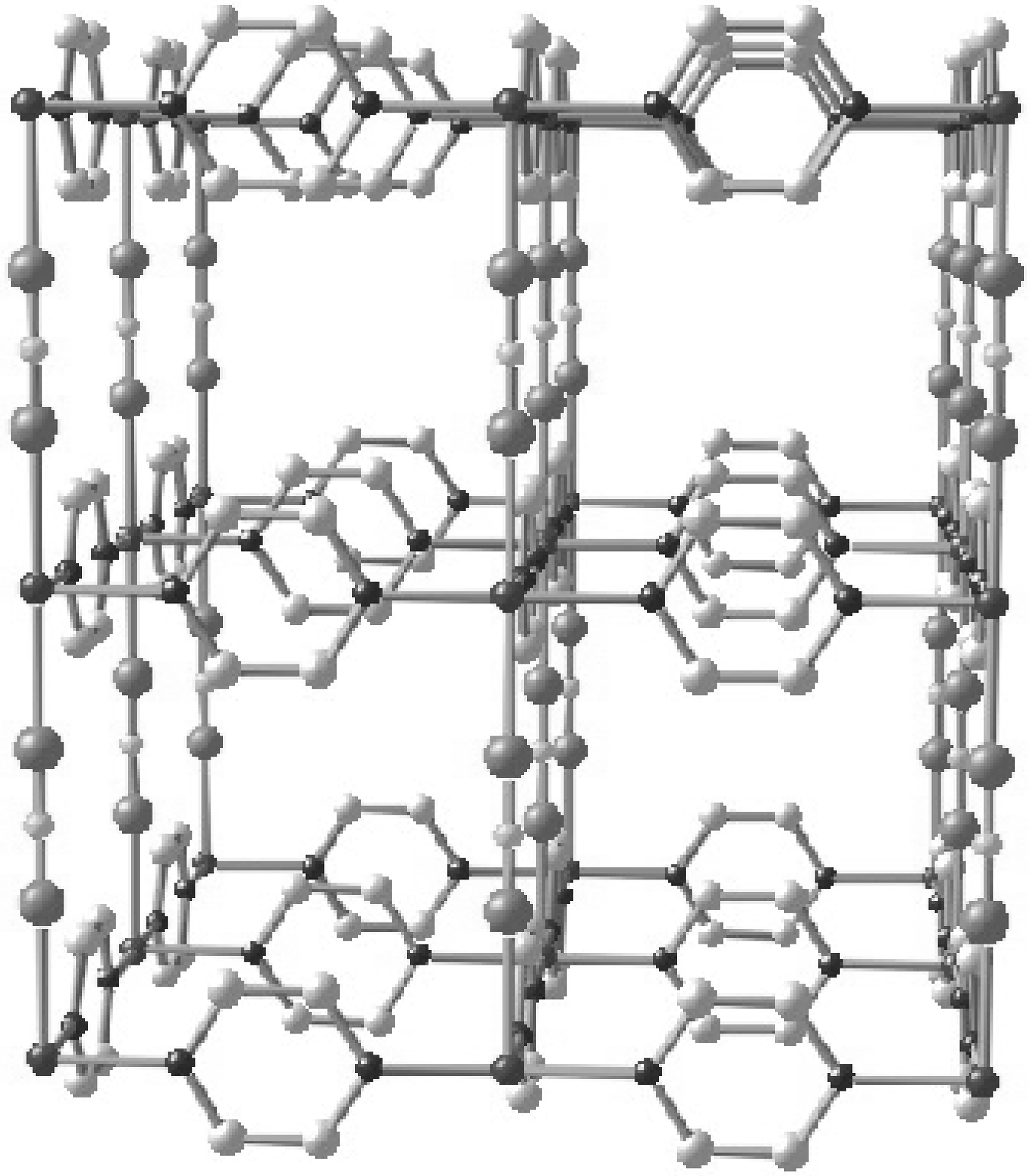}
\includegraphics[width=6cm]{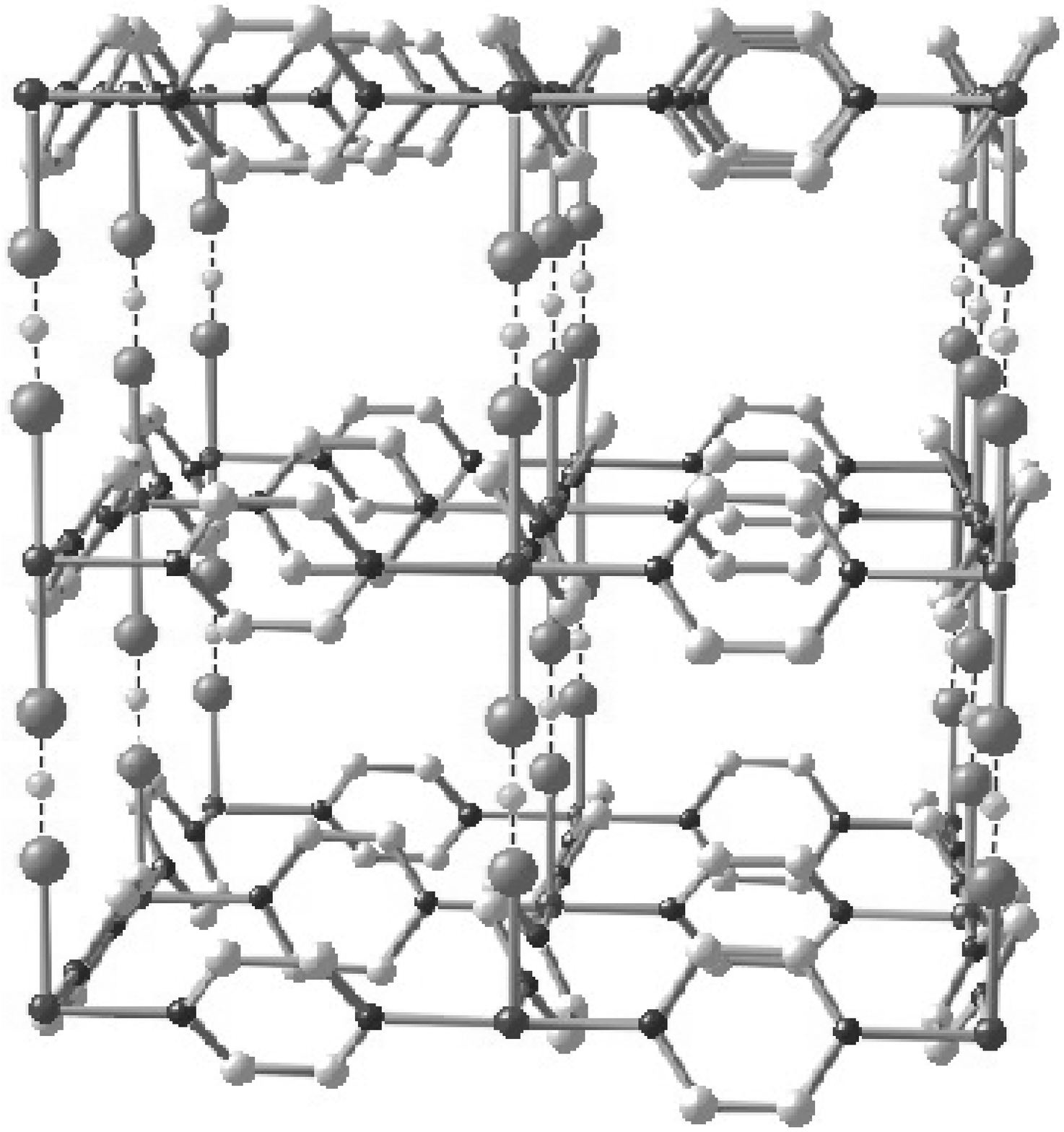}
\caption{Experimentally-determined~\cite{mansonchemcomm,growth} 
room-temperature crystal structures of
[Cu(HF$_2$)(pyz)$_2$]X 2D organic magnets.
In each figure, the Cu-F-H-F-Cu linkages
(parallel to $c$ and perpendicular to the
$ac$ planes) are vertical.
The upper figure is for X = SbF$_6$
whilst the lower shows
X = BF$_4$.
Note that for X = BF$_4$ (lower), the planes of the pyrazine
molecules are tilted by $31.6^\circ$ away from
being orthogonal to the $ab$ planes.
However, for X = SbF$_6$ (upper), the 
planes of the pyrazine molecules are perpendicular to
the $ab$ planes.
(The [SbF$_6$]$^-$ and [BF$_4$]$^-$ anions
have been omitted for clarity;
they inhabit the ``body-center''
sites of each approximately cubic unit.)}
\label{fig4}
\end{figure}

Having established a reliable method for deriving $J$ and the anisotropy
from $M(B)$ and $T_{\rm N}$,
we focus our remaining discussion on the compounds with formula 
[Cu(HF$_2$)(pyz)$_2$]X,~\cite{mansonchemcomm,growth} where X can be
ClO$_4$, BF$_4$, PF$_6$ {\it etc.} (Table~\ref{table1}). 
All of these materials
possess very similar extended polymeric structures consisting of
2D, four-fold symmetric [Cu(pyz)$_2$]$^{2+}$ sheets in
the $ab$-planes that are connected along the
$c$-axis by linearly-bridging [HF$_2$]$^-$ ions
(Fig.~\ref{fig4})~\cite{mansonchemcomm,growth}.
The Cu-Cu separations are similar
along the Cu-(pyz)-Cu (0.6852~nm) and Cu-FHF-Cu
(0.6619~nm) linkages, so that the structure
may be described as 
pseudo-cubic~\cite{mansonchemcomm,growth};
the X anions are placed in the body-centre positions
within each ``cube''. The Cu-F and Cu-(pyz)
bonds result in the Cu$^{2+}$ $d_{x^2-y^2}$
orbitals lying within the $ab$ planes,
as evidenced by the $g$-factor anisotropy observed
in EPR measurements~\cite{cox}; as noted above,
the $ab$ planes
also correspond to the 2D planes within
which the strong exchange pathways
occur.

There is little variation
of the lattice parameters across the 
[Cu(HF$_2$)(pyz)$_2$]X 
series~\cite{mansonchemcomm,growth};
given this similarity,
it is at first sight surprising that
the exchange parameters $J$ in Table~\ref{table1}
fall into two distinct groups: the 
compounds with tetrahedral anions
(X = BF$_4$ and ClO$_4$) possess
values $J\approx -7$~K and those with octahedral
anions (X = PF$_6$, SbF$_6$ and AsF$_6$)
have $J\approx -13$~K.
Note also that the compounds with tetrahedral anions
are more anisotropic ($|J_{\perp}/J|\sim 10^{-3}$)
than those with octahedral anions ($|J_{\perp}/J|\sim 10^{-2}$).

We first discuss whether
the X anions can play a direct role
as exchange pathways between the
Cu$^{2+}$ ions.
First of all, the anions
[ClO$_4$]$^-$ and [BF$_4$]$^-$ have
radically different electronic orbitals, and yet the
intralayer exchange energies for the
magnets containing these anions
are very similar (see the first three rows of Table~\ref{table1}).
Moreover, as mentioned above, the X anions are not 
within the 2D Cu$^{2+}$ planes, but at
body-center positions within the
``cubes'' (Fig.~\ref{fig4}).
The Cu-(anion) separation
is therefore roughly the same for
the interlayer and intralayer directions;
if the anions played a direct role as exchange
pathways, one might expect that the  
[Cu(HF$_2$)(pyz)$_2$]X compounds would
have $|J_{\perp}/J|$ values that were
somewhat larger (i.e. more isotropic)
than the values $\sim 10^{-3}-10^{-2}$
that are actually observed (see Table~\ref{table1}).
Therefore, instead of playing a direct role in the
exchange, it is more likely that it is
an anion {\it size effect} on the crystal structure
that is affecting the exchange pathways.

The structural difference that 
may well be responsible
for the factor of two 
change in $J$
is the configuration of the pyrazine
molecules within the Cu-(pyz)-Cu linkages.
Fig.~\ref{fig4} (upper) shows
[Cu(HF$_2$)(pyz)$_2$]SbF$_6$,
one of the systems with octahedral anions;
the structures of X = AsF$_6$, PF$_6$ are very similar.
At room temperature,
the larger size of the octahedral anions
compared to the tetrahedral examples
forces the pyrazine ligands
to stand up virtually
perpendicular to the $ab$ planes~\cite{growth} (Fig.~\ref{fig4}).
By contrast, the planes of the
pyrazine ligands in the compounds with the
smaller tetrahedral anions, [Cu(HF)$_2$(pyz)$_2$]BF$_4$
(Fig.~\ref{fig4} (lower)) and 
[Cu(HF)$_2$(pyz)$_2$]ClO$_4$, are not 
perpendicular to the $ab$ plane, but
are tilted away by $31.6^\circ$ (X=BF$_4$)
or $25.8^\circ$ (X=ClO$_4$)
in a pattern that preserves the four-fold symmetry
of the Cu$^{2+}$ sites~\cite{growth}.
As one lowers the temperature~\cite{growth},
the tilt of the pyrazine ligands increases
in all of the compounds 
(i.e. those with octahedral and tetrahedral
anions); however, in the case of the
octahedral anions, space-filling considerations
restrict the tilt to values in the range $5^\circ$
(X = PF$_6$) to $12^\circ$ (X = AsF$_6$).
This suggests that it
may be the orientation dependence of the
pyrazine ligand that produces the
factor two difference in $J$, with
the more perpendicularly-disposed pyrazines
(Fig.~\ref{fig4}) presenting a more
efficient exchange pathway within the layers.

\section{Summary}
Magnetization experiments
have been carried out on nine organic
Cu-based 2D 
Heisenberg magnets. 
These systems exhibit a low-$T$
magnetization that is concave as a function of field,
with a sharp ``elbow'' transition 
to a constant saturation value at a critical field $B_{\rm c}$.
Monte-Carlo simulations including interlayer exchange
quantitatively reproduce the 
data; the concavity indicates the effective dimensionality 
and $B_{\rm c}$ is an accurate measure of the
in-plane exchange energy $J$. 
Taken in conjunction with Ne\'el temperatures derived from
muon-spin rotation,
the values of $J$ may be used to obtain quantitative estimates 
of the exchange anisotropy, $|J_{\perp}/J|$.

We suggest that 
in organic magnets of the form [Cu(HF$_2$)(pyz)$_2$]X,
where X is an anion molecule,
the sizes of $J$ and $J_{\perp}$ may be controlled
by the tilting of the pyrazine 
molecule with respect to the 2D planes.
Thus, it may be possible
to use molecular architecture to design
magnets with very specific values
of $J$, tailored to a particular desired property.
One possible application would be an organic
magnet designed to simulate the antiferromagnetic
interactions germane to cuprate 
superconductivity~\cite{manousakis,schrieffer,dai,stock,julian,harrison},
but with exchange energies small enough to permit manipulation
of the magnetic groundstate using standard laboratory fields.

\section*{Acknowledgements}
We thank Peter Baker
and Alex Amato for experimental assistance
and Bill Hayes for stimulating discussions.
This work is supported by the
US Department of Energy (DoE) BES
program ``Science in 100~T''.
Work at NHMFL occurs under
the auspices
of the National Science
Foundation, DoE and the State of Florida. 
Work at Argonne is
supported by the Office of Basic Energy Sciences,
DoE (contract DE-AC02-06CH11357).
Part of this work was carried out at the
Swiss Muon Source, Paul Scherrer Institut,
CH, and at the ISIS Facility, Rutherford-Appleton
Laboratory, UK.
Research at EWU was supported by an award from the 
Research Corporation. 
TL and
PAG acknowledge support from
the Royal Commission of the Exhibition of
1851 and the Glasstone Foundation respectively.
JS thanks Oxford University for the provision of
a Visiting Professorship.


\begin{thebibliography}{00}
\bibitem{q2dreview}
M.A. Kastner, R.J.~Birgenau, G.~Shirane amd Y.~Endoh, 
Rev. Mod. Phys. {\bf 70},
897 (1998).
\bibitem{schollwock}
U.~Schollw\"{o}ck, D.J.J.~Farnell and R.F.~Bishop (eds),
{\it Quantum magnetism}, (Springer, Berlin, 2004).
\bibitem{manousakis}
E. Manousakis, Rev. Mod. Phys. {\bf 53}, 1 (1991).
\bibitem{christensen}
N.B.~Christensen, H.M.~Ronnow, D.F.~McMorrow,
A.~Harrison, T.G.~Perring, M.~Enderle, R.~Coldea,
L.P.~Regnault and G.~Aeppli, Proc. Nat. Acad. Sci.
{\bf 104}, 15264 (2007).
\bibitem{vajk}
O.P.~Vajk, P.K.~Mang, M.~Greven, P.M.~Gehring and J.W.~Lynn,
Science {\bf 295}, 1691 (2002).
\bibitem{deng}
D.S.~Deng, X.F.~Jin and R.~Tao, Phys. Rev. B {\bf 65},
132406 (2002).
\bibitem{beard}
B.B.~Beard, A.~Cuccoli, R. Vaia and P. Verrucchi,
Phys. Rev. B {\bf 68}, 104406 (2003).
\bibitem{sengupta}
P. Sengupta, A.V. Sandvik and and R.R.P.~Singh, 
Phys. Rev. B {\bf 68}, 094423 (2003).
\bibitem{zhang}
R. Zhang and S.~Zhu, Phys. Lett. A {\bf 348}, 110 (2006).
\bibitem{schrieffer}
J.R.~Schrieffer and J.S.~Brooks (eds),
{\it High temperature superconductivity theory and
experiment} (Springer, Berlin, 2007).
\bibitem{dai}
P.~Dai, H.A. Mook, R.D. Hunt and F.~Do\u{g}an,
Phys. Rev. B {\bf 63}, 054525 (2001).
\bibitem{stock}
C.~Stock, W.J.L. Buyers, R.A. Cowley, P.S. Clegg, R. Coldea,
C.D. Frost, R. Liang, D. Peets, D. Bonn, W.N. Hardy and R.J. Birgenau,
Phys. Rev. B {\bf 71}, 024522 (2005).
\bibitem{julian}
S.R. Julian and M. Norman,
Nature {\bf 447}, 537 (2007).
\bibitem{harrison}
N.Harrison, R.D.~McDonald and J.~Singleton, 
preprint arXiv:0710.1932 (2007).
\bibitem{pines}
P.Monthoux, D.~Pines and G.G.~Lonzarich,
Nature {\bf 450}, 1177 (2007).
\bibitem{mermin}
N.D. Mermin and H. Wagner, Phys. Rev. Lett.
{\bf 17}, 1133 (1966).
\bibitem{choi}
J.~Choi, J.D. Woodward, J.L. Musfeldt,
C.P. Landee and M.M. Turnbull,
Chem. of Materials {\bf 15}, 2797 (2003).
\bibitem{mansonchemcomm}
J.L. Manson, M.M.~Conner, J.A.~Schlueter,
T.~Lancaster, S.J.~Blundell, M.L.~Brooks,
T.~Papageorgiou, A.D.~Bianchi, J. Wosnitza
and M.H.~Wangbo, Chem. Comm. {\bf 2006}, 4894 (2006).
\bibitem{lancasterprb}
T. Lancaster, S.J.~Blundell, M.L. Brooks, P.J.~Baker,
F.L.~Pratt, J.L.~Manson, M.M.~Conner, F.~Xiao, C.P.~Landee,
F.A.~Chaves, S.~Soriano, M.A.~Novak, T.P.~Papageorgiou, 
A.D.~Bianchi, T.~Herrmannsd\"{o}rfer, J.~Wosnitza and J.A.~Schlueter,
Phys. Rev. B {\bf 75},
094421 (2007).
\bibitem{deumel}
M.~Deumel, C.P.~Landee, J.J.~Novoa, M.A.~Robb and
M.M.~Turnbull, Polyhedron {\bf 22}, 2235 (2003).
\bibitem{lancaster1D}
T. Lancaster, S.J.~Blundell, M.L.~Brooks, P.J. Baker,
F.L~Pratt, J.L.~Manson, C.P.~Landee and C.~Baines, 
Phys. Rev. B {\bf 73},
020410 (2006).
\bibitem{blundell}
S.J. Blundell, T.~Lancaster, F.L.~Pratt
P.J.~Baker, M.L.~Brooks, C.~Baines, J.L.~Manson and C.P.~Landee, 
J. Phys. Chem. Solids, {\bf 68},
2039 (2007). 
\bibitem{growth} 
J.L. Manson, H. Southerland, J. Schlueter and
K. Funk, preprint (2008).
\bibitem{entangle}
T. Lancaster, S.J.~Blundell, P.J.~Baker,
M.L.~Brooks, W.~Hayes, F.L.~Pratt, J.L.~Manson
M.M.~Conner and J.A.~Schlueter, Phys. Rev. Lett.
{\bf 99}, 267601 (2007).
\bibitem{cox}
S. Cox, R.D.~McDonald, K. Funk, H.A.~Southerland,
J.L.~Manson and J.A.~Schlueter, preprint (2008).
\bibitem{abragam}
A. Abragam et B. Bleaney,
{\it R\'esonance paramagn\'etique \'electronique
des ions de transition} (Presses Universitaires de France,
Paris, 1971), p 208.
\bibitem{goddard}
P.A. Goddard, J.~Singleton, A.L.~Lima-Sharma,
E.~Morosan, S.J.~Blundell, S.L.~Bud'ko and P.C.~Canfield,
Phys. Rev. B., {\bf 75}, 094426 (2007).
\bibitem{ho}
P.-C. Ho, J.~Singleton, M.B.~Maple, H.~Harima,
P.A.~Goddard, Z.~Henkie amd A.~Pietraszko, 
New J. Physics,
{\bf 9}, 269 (2007).
\bibitem{boebinger}
M.~Jaime, A. Lacerda, Y. Takano and G.S. Boebinger,
Institute of Physics: Conference Series
{\bf 51}, 643 (2006).
\bibitem{sse1}
A. W. Sandvik and J. Kurkij\"arvi, 
Phys. Rev. B {\bf 43}, 5950 (1991);
A. W. Sandvik, {\it ibid.} 
{\bf 56}, 11678 (1997).
\bibitem{sse2}
A. W. Sandvik, Phys. Rev. B {\bf 59}, R14157 (1999).
\bibitem{dloops}
O. F. Sylju{\aa}sen and \
A. W. Sandvik, Phys. Rev. E {\bf 66}, 046701 (2002).
\bibitem{tempering1}
E. Marinari, Lecture Notes in Physics, 
Vol.~501 {\it Advances in computer
simulation: lectures held at the 
E\"{o}tv\"{o}s Summer School in Budapest,
Hungary, 16-20, July 1996}, 
edited by J. Kertsz and I. Kondor (Springer, 1998).
\bibitem{tempering2}
K. Hukushima, H. Takayama, K. Nemoto, 
Int. J. Mod. Phys. C {\bf 7},
337 (1996); K. Hukushima, K. Nemoto, J. Phys. Soc. Jpn. 
{\bf 65}, 1604 (1996)
\bibitem{bow}
P. Sengupta, A. W. Sandvik, and D. K. Campbell,
Phys. Rev. B 65, 155113 (2002).
\bibitem{rosspoint}
From Eq.~\ref{eq1} it is easy to see 
that $g\mu_{\rm B}B_{\rm c}=4J+2J_{\perp}$.
However, without the information on the
anisotropy given by the curvature of
the $M(B)$ data, this expression contains
two unknowns. Moreover, it was found that the
best method (i.e. that with the
smallest uncertainties and that best utilized all of the data)
for determining $B_{\rm c}$ was that described in the
text; matching the $M(B)$ curves to the Monte Carlo
simulations by varying $B_{\rm c}$.
\bibitem{landee}
C.P. Landee, S.A. Roberts and R.D. Willett, 
J. Chem. Phys. {\bf 68}, 4574 (1978).
\bibitem{footnote}
The definition of $J$ used in
Ref.~\cite{landee} differs by a factor 2
from the one employed in the present paper.
\bibitem{yasuda}
C. Yasuda, S.~Todo, K.~Hukushima, A.~Alet, M.~Keller
M.~Troyer and H.~Takayama,
Phys. Rev. Lett. {\bf 94}, 217201 (2005).
\end{thebibliography}
\end{document}